\documentclass[epj]{svjour}
\usepackage{graphics}

\newcommand{\ba}{\begin{array}}
\newcommand{\ea}{\end{array}}
\newcommand{\be}{\begin{equation}}
\newcommand{\ee}{\end{equation}}
\newcommand{\bea}{\begin{eqnarray}}
\newcommand{\eea}{\end{eqnarray}}
\newcommand{\beq}{

\begin{equation}}
\newcommand{\eeq}{\end{equation}}

\begin{document}
\title{Updated NLL Results for $\bar B \rightarrow X_{s,d} \gamma$ in and beyond the SM\thanks{Contribution to the International Europhysics Conference on High Energy Physics EPS03, 17-23 July 2003, Aachen,  \mbox{Germany}, presented by T.H.}}
\author{T.~Hurth\thanks{{Heisenberg Fellow}}\inst{1}
\and E.~Lunghi\inst{2}  \and W.~Porod \inst{2}}

\institute{Theoretical Physics Division, CERN, CH-1211 Geneva 23, Switzerland
 and\\ SLAC, Stanford University, Stanford, CA 94309, USA \and 
Institute for Theoretical Physics, University of Zurich, CH-8057 
 Zurich, Switzerland}

\date{{\tt CERN-TH/2003-251}, {\tt SLAC-PUB-10210}, {\tt ZU-TH 17/03}}

\abstract{  
We present  general model-independent formulae
for the branching ratios and the direct tagged CP asymmetries 
for the  inclusive $\bar B \rightarrow X_d\, \gamma$ and 
 $\bar B \rightarrow X_s\,  \gamma$ modes. We also  
 update the corresponding SM predictions.}


\authorrunning{T.Hurth, E.Lunghi, W.Porod}
\titlerunning{Updated NLL Results for $\bar B \rightarrow X_{s,d} \gamma$ 
in and beyond the SM}

\maketitle

\section{Introduction}
\label{Intro}
In the near future more precise data on the inclusive decay 
$B \rightarrow X_s \gamma$ is expected from the $B$ factories, 
but also the present experimental accuracy already reached the $10 \%$ 
level as reflected \cite{aleph,belle,cleobsg,babar1,babar2,jessop}
in the world average of the present measurements:
\beq
\label{world}
{\cal B}(\bar B \to X_s \gamma) = (3.34 \pm 0.38) \times 10^{-4}.  
\eeq
In addition, direct CP asymmetries within this mode are now within 
experimental reach~\cite{cleoCP,belleCP}:
\bea 
A_{\rm CP} (\bar B \to X_s \gamma) & = & 
\cases{
-0.079 \pm 0.108_{\rm stat} \pm 0.022_{\rm syst} \cr 
-0.004 \pm 0.051_{\rm stat} \pm 0.038_{\rm syst} \cr} 
\eea
In the first measurement of  CLEO there is a small contamination 
of the $\bar B \rightarrow X_d \gamma$ mode.

All these measurements are compatible with the standard model (SM) 
predictions and thus lead to severe constraints on new physics 
models \cite{Carena,Degrassinew,Giannew,OUR,NEWNEW,Luca}, which 
represents
very valuable information for the direct search for physics
beyond  the SM (for recent reviews,  
see~\cite{hurth,yellowbook,durham}).

A direct 
measurement of the inclusive $\bar B \rightarrow X_d \gamma$ 
mode is rather difficult,  but perhaps still within the reach of the 
present high-luminosity $B$ factories. However, the CP violation
within that mode can be perhaps tested indirectly by an untagged 
CP measurement (see below).

In this letter  we present general model-independent formulae 
for the  branching ratios and the direct tagged CP asymmetries 
for the inclusive $\bar B \rightarrow X_{s,d} \gamma$ modes as 
explicit numerical expressions for these
observables as functions of Wilson coefficients and CKM angles. The
extraction of the latter from experimental data depends critically on
the assumptions about the presence and the structure of new physics
contributions to several key observables. 

For this purpose we update and generalize the SM results at NLL level 
given  in Refs. \cite{GM,Buras} and \cite{AG7,KaganNeubert,SoniWu} 
in order to accommodate new physics models with new CP-violating phases
and also implement several improvements.
For a detailed discussion of our results we refer the reader 
to a forthcoming paper \cite{NewCP}.

\section{NLL Predictions}

The general effective hamiltonian that governs 
the inclusive $\bar B\to X_q \gamma$ decays ($q=d,s$) {in the SM} 
is
\bea
H_{\rm eff} (b\rightarrow q \gamma) &=&
- {4 G_F\over \sqrt{2}} V_{tb}^{} V_{tq}^* \times  \\ 
&\times& \left(\sum_{i=1}^8 C_i  {\cal O}_i  +  \epsilon_q \, \sum_{i=1}^2 
C_i  ({\cal O}_i - {\cal O}_i^u ) \right) \nonumber
\label{bqgEH}
\eea
where $\epsilon_q = (V_{ub}^{} V_{uq}^*) / (V_{tb}^{} V_{tq}^*)$ and
the most relevant operators are:
\bea
\label{oper}
   {\cal O}_1 & = & (\bar{q}_{L} \gamma_\mu T^a c_{ L}) (\bar{c}_{ L} 
                  \gamma^\mu T^a b_{ L}),\nonumber  \\
   {\cal O}_1^u & = & (\bar{q}_{L} \gamma_\mu T^a u_{ L}) (\bar{u}_{ L} 
                   \gamma^\mu T^a b_{ L}), \nonumber \\
   {\cal O}_2    & = & (\bar{s}_{L}\gamma_{\mu}  c_{L })
                (\bar{c}_{L }\gamma^{\mu} b_{L}) \, , \nonumber \\
    {\cal O}^u_2    & = & (\bar{s}_{L}\gamma_{\mu}  u_{L })
                (\bar{u}_{L }\gamma^{\mu} b_{L}) \, , \nonumber \\
    {\cal O}_7    & = & ({e}/{16 \pi^2}) m_b (\bar{s}_{L} \sigma^{\mu\nu}
                b_{R}) F_{\mu\nu} \,  \nonumber \\
    {\cal O}_8    & = & ({g_s}/{16 \pi^2}) m_b (\bar{s}_{L} \sigma^{\mu\nu}
                T^a b_{R}) G_{\mu\nu}^a\, . \nonumber  
\eea
The subscripts $L$ and $R$ refer to left- and right-handed
components of the fermion fields. In $b\to s$ transitions the
contributions proportional to $\epsilon_s$ are rather small, while in
$b\to d$ decays the $\epsilon_d$ term is of the same order as 
the first term in effective hamiltonian.  


Regarding the input parameters we focus here on the 
 issue of the charm mass definition in the
matrix element of ${\cal O}_2$:
In Ref.~\cite{GM}, it is argued that all the factors 
of $m_c$ come from propagators
corresponding to charm quarks that are off-shell by an amount $\mu^2
\sim m_b^2$. It seems, therefore, more reasonable to use the
$\overline{\rm MS}$ running charm mass at a scale $\mu$ in the range
$(m_c,m_b)$. The reference values of the
charm and bottom masses are $m_c = m_c^{\rm \overline MS} (m_c^{\rm
  \overline MS}) = (1.25\pm 0.10)\, {\rm GeV}$ and $m_b = m_b^{1S}$, 
where 
the  $1S$ mass of the $b$ quark is defined as half of the
  perturbative contribution to the $\Upsilon$ mass as usual:
$m_b^{1S} = (4.69 \pm 0.03 )\, {\rm GeV}$.  We
first fix the central value of $m_c = 1.25\, {\rm GeV}$ 
and vary $\mu$; then
we add in quadrature the error on $m_c$ ($\delta_{m_c} = 8\%$). The
resulting determination is:
\beq
{m_c \over m_b} = 0.23 \pm 0.05 \, .
\label{myz}
\eeq
The pole mass choice corresponds, on the other hand, to ${m_c \over
  m_b} = 0.29 \pm 0.02$. Note that the question whether to use the
running or the pole mass is, strictly speaking, a NNLL issue. The most
conservative position consists in accepting any value of $m_c/m_b$
that is compatible with any of these two determinations: $0.18 \leq
m_c/m_b \leq 0.31$. Taking into account our experience on higher-loop 
computations,  we are  led to the educated guess 
that the central value $m_c/m_b =0.23$ represents the
best possible choice, but we allow for a large asymmetric error that
fully covers the above range (and that reminds us of this problem that
can be solved only via a NNLL computation):
\beq
{m_c \over m_b} = 0.23_{-0.05}^{+0.08} \, .
\label{ourz}
\eeq

We present our SM updates for two different energy
cuts within the photon spectrum $E_0 = (1.6 \, {\rm GeV} , {m_b / 20})$.
There are four sources of uncertainties: the charm mass
($\delta_{m_c/m_b}$), the CKM factors ($\delta_{\rm CKM}(s) = 0.5\%$,
$\delta_{\rm CKM}(d) = 11\%$), the parametric uncertainty, 
including that of the overall normalization,
$\alpha_s$ and $m_t$ ($\delta_{\rm param}$),  and 
the perturbative scale uncertainty ($\delta_{\rm scale}$):
\bea 
{\cal B}(\bar B \to X_s \gamma;\, E_\gamma > 1.6 \, {\rm GeV} ) \times 10^{4} &=& 
\nonumber \\
&& \hskip -6.4cm (3.56 \,\, \left. {}^{+0.24}_{-0.40} \right|_{m_c \over m_b}
                     \pm 0.02_{\rm CKM} \pm 0.24_{\rm param.} \pm 0.14_{\rm scale}) \\
{\cal B}(\bar B \to X_d \gamma;\, E_\gamma > 1.6 \, {\rm GeV} ) \times 10^{5} &=& \nonumber \\
&& \hskip -6.4cm  (1.36 \,\,  \left. {}^{+0.14}_{-0.21}   \right|_{m_c \over m_b} \pm 0.15_{\rm CKM} \pm 0.09 _{\rm param.} \pm 0.05_{\rm scale}) \\
{\cal B}(\bar B \to X_s \gamma;\,  E_\gamma > m_b/20 ) \times 10^{4} &=& \nonumber \\
&& \hskip -6.4cm (3.74 \,\,  \left. {}^{+0.26}_{-0.44}   \right|_{m_c \over m_b} 
                     \pm 0.02_{\rm CKM} \pm 0.25_{\rm param.} \pm 0.15_{\rm scale}) \\
{\cal B}(\bar B \to X_d \gamma; E_\gamma > m_b/20 ) \times 10^{5} &=& \nonumber \\
&& \hskip -6.4cm (1.44 \,\,  \left. {}^{+0.15}_{-0.23}   \right|_{m_c \over m_b}
                     \pm 0.16_{\rm CKM} \pm 0.10_{\rm param.} \pm 0.06_{\rm scale}). 
\eea
The  CKM uncertainties are almost negligible in $b\to s \gamma$ transitions 
but play an important role in $b\to d \gamma$ ones. This implies the large 
impact on the CKM phenomenology of the latter. 
We note that in the $b \rightarrow d$ mode there is an additional
uncertainty due to the $up$ quark loops which is suppressed
by $\Lambda_{QCD}/m_b$ (for details see \cite{hurth}).

The direct CP asymmetries in $\bar B \to X_q \gamma$ are  defined by 
\bea
A_{CP}^{b\to q \gamma} & \equiv & { \Gamma{[\bar B \to X_q \gamma]} - \Gamma{[B \to X_{\bar q} \gamma]} \over
                                    \Gamma{[\bar B \to X_q \gamma]} + \Gamma{[B \to X_{\bar q} \gamma]}}. 
\label{acp1}
\eea
It was shown that the CP asymmetry in the $b \rightarrow s$ mode is below
$1 \%$ \cite{AG7,KaganNeubert,SoniWu} within the SM. 
This small value is a result of
three suppression factors.  There is an $\alpha_s$ factor needed in
order to have a strong phase; moreover, there is a CKM suppression of
order $\lambda^2$ and there is a GIM suppression of order
$(m_c/m_b)^2$,  reflecting the fact that in the limit $m_c = m_u$ any CP
asymmetry in the SM would vanish.
Within the SM the CP asymmetry in the $b \rightarrow d$ mode is enhanced,  
 with respect to  the one in the $b \rightarrow s$ mode,  by the CKM  
factor $[\lambda^2 \, ((1-\rho)^2 + \eta^2)]^{-1}$.

We update the SM predictions,  which are essentially independent of the 
photon energy cut-off ($E_0$) and get (for $E_0=1.6 \, {\rm GeV}$):
\bea 
A_{\rm CP}^{b \to s \gamma}&=& 
( 0.44 \, \left. {}^{+0.15}_{-0.10} \right|_{m_c \over m_b}  
                \left. \pm 0.03_{\rm CKM} {}^{+0.19}_{-0.09} 
\right|_{\rm scale}) \%\\ 
A_{\rm CP}^{b \to d \gamma}&=& 
(-10.2 \, \left. {}^{+2.4}_{-3.7} \right|_{m_c \over m_b}  
                \left.  \pm 1.0_{\rm CKM}   {}^{+2.1}_{-4.4}  \right|_{\rm scale}) \%.
\eea
The additional parametric uncertainties are subdominant. However, 
the scale uncertainties are rather large because the CP asymmetries arise 
at the $O(\alpha_s)$ only.  This purely perturbative uncertainty can be 
removed by a NNLL QCD calculation. 

The so-called untagged CP asymmetry 
$A_{\rm CP}^{b \, \to \,(s+d)\, \gamma}$
is the favoured observable, at least  from the theoretical point of view.  
A simple expression of this observable is given by 
\bea
A_{\rm CP}^{b \to (s+d) \gamma} &=& 
{A_{\rm CP}^{b \to s \gamma} + R_{ds} \; A_{\rm CP}^{b \to d \gamma}
     \over 1 + R_{ds}} \, ,
\label{untag}
\eea
$R_{ds}=\Sigma\Gamma_d/\Sigma\Gamma_s$,\, $\Sigma\Gamma_q :=\Gamma(\bar B\to X_q \gamma)+\Gamma (B \to X_{\bar q} \gamma)$.

As was first noticed in \cite{Soares}, 
the untagged CP asymmetry vanishes within the SM if the
U-spin limit is considered. This is a direct consequence of CKM unitarity.
Within the inclusive channels, one can rely on parton--hadron duality
and can actually compute the U-spin breaking by keeping a
non-vanishing strange quark mass \cite{mannelhurth}.  
In \cite{mannelhurth2} U-spin breaking effects were estimated and found 
to be completely negligible, even beyond the leading partonic contribution 
within  the heavy mass expansion. 
Thus, the measurement of the untagged CP asymmetry provides 
a very clean SM test,
whether generic new CP phases are active or not.  Any significant
deviation from the SM zero prediction would be a direct hint of  non-CKM
contributions to CP violation. An analysis of the untagged asymmetry within
various new physics scenarios will be presented in \cite{NewCP}.

\section{Model-independent Formulae}
\label{NewPhysics}

We assume within our model-independent analysis of new physics effects 
that  the dominat ones  only modify the
Wilson coefficients of the dipole operators ${\cal O}_7$ and ${\cal O}_8$ 
and also 
introduce contributions proportional to the corresponding operators with 
opposite chirality:
\bea
{\cal O}_7^R &=&({e }/{16 \pi^2})\, m_b  (\bar{q}_{R} \sigma_{\mu \nu} b_{L}) F^{\mu \nu},\\
{\cal O}_8^R & =& ({g_s }/{16 \pi^2})\, m_b  (\bar{q}_{R} T^a \sigma_{\mu \nu} b_{L}) G^{a \mu \nu} \; .
\label{operatorchiral}
\eea
This is known as a very good approximation for the most relevant 
new physics scenarios.

Within our model-independent formulae for the branching ratios and CP
asymmetries, the Wilson coefficients $C_{7,8(R)}$ and $C_{7,8}$ and
all the CKM ratios are left unspecified.  The explicit derivation of
the formulae given below can be found in \cite{NewCP}.  The branching
ratio can be written as
\bea
{\cal B}(\bar B \to X_q \gamma)
& = & 
{{\cal N}\over 100}   \left| V_{tq}^* V_{tb}^{}\over V_{cb}^{}\right|^2  \, 
{\cal B}^{\rm unn} \,,
\eea
where ${\cal N} = 2.567 \, (1 \pm 0.064 ) \times 10^{-3}$ is an
overall normalization factor, the ratios $R_{7,8}$ and $\widetilde
R_{7,8}$ are
\bea
R_{7,8} = { (C_{7,8}^{(0) {\rm SM}}+C_{7,8}^{(0) {\rm NP}})(\mu_0)
\over C_{7,8}^{(0) {\rm SM}}(m_t)},\,\, 
\widetilde R_{7,8} = {C_{7,8 R}^{(0) {\rm NP}}(\mu_0) \over C_{7,8}^{(0) {\rm SM}}(m_t)}, \nonumber
\eea
and the unnormalized branching ratio is
\bea
{\cal B}^{\rm unn} & = & 
\Big[
a  + a_{77} \, (|R_7|^2 + |\widetilde R_7|^2) + a_7^r \, {\rm Re} (R_7)   + a_7^i \, {\rm Im} (R_7)  
\nonumber \\ 
& &  \hskip -1.2cm 
+ a_{88} \, (|R_8|^2+ |\widetilde R_8|^2) + a_8^r \, {\rm Re} (R_8)  + a_8^i \, {\rm Im} (R_8)
\nonumber \\ 
& & \hskip -1.2cm 
+ a_{\epsilon\epsilon} \, |\epsilon_q|^2 + a_\epsilon^r \, {\rm Re} (\epsilon_q)  
+ a_\epsilon^i \, {\rm Im} (\epsilon_q) 
+ a_{87}^r \, {\rm Re} (R_8^{} R_7^* + \widetilde R_8^{} \widetilde R_7^*) 
\nonumber \\
& & \hskip -1.2cm 
+ a_{7\epsilon}^r \, {\rm Re} (R_7^{} \epsilon_q^*) 
+ a_{8\epsilon}^r\, {\rm Re} (R_8^{} \epsilon_q^*) 
+ a_{87}^i \, {\rm Im} (R_8^{} R_7^*+ \widetilde R_8^{} \widetilde R_7^*) 
\nonumber \\
& &  \hskip -1.2cm 
+ a_{7\epsilon}^i \, {\rm Im} (R_7^{} \epsilon_q^*) + a_{8\epsilon}^i \, {\rm Im} (R_8^{} \epsilon_q^*) 
\Big] \, . 
\label{brnum}
\eea
The CP asymmetry is given by
\bea
A_{\rm CP}^{b\to q \gamma}
& = & {1 \over {\cal B}^{\rm unn}} \; 
{\rm Im} 
\big[   
a_7^i \, R_7 + a_8^i \, R_8 + a_\epsilon^i \, \epsilon_q 
\nonumber \\ 
& & \hskip -1cm
+ a_{87}^i \, (R_8^{} R_7^*   + \widetilde R_8^{} \widetilde R_7^*) 
+ a_{7\epsilon}^i \, R_7^{} \epsilon_q^* + a_{8\epsilon}^i \, 
R_8^{} \epsilon_q^*  \big] \,.
\label{acpnum}
\eea
The numerical values of the coefficient functions are collected in
Table~\ref{tab:coeff}.

\begin{table}
\begin{center}
\begin{tabular}{|c|cc|cc|} \hline
  & \multicolumn{4}{c|}{NLL}\cr \hline
 $E_0  $ & \multicolumn{2}{c|}{$1.6 \, \hbox{GeV}$} & \multicolumn{2}{c|}{$m_b/20$} \cr
 $m_c/m_b  $ & 0.23 &0.29 & 0.23 & 0.29 \cr \hline\hline
 $ a $ & 7.8221& 6.9120& 8.1819& 7.1714\cr 
 $ a_{77}$& 0.8161& 0.8161& 0.8283& 0.8283\cr 
 $ a_7^r $& 4.8802& 4.5689& 4.9228& 4.6035\cr 
 $ a_7^i $& 0.3546& 0.2167& 0.3322& 0.2029\cr 
 $ a_{88}$& 0.0197& 0.0197& 0.0986& 0.0986\cr 
 $ a_8^r $& 0.5680& 0.5463& 0.7810& 0.7600\cr 
 $ a_8^i $&-0.0987&-0.1105&-0.0963&-0.1091\cr 
 $ a_{\epsilon\epsilon}$& 0.4384& 0.3787& 0.8598& 0.7097\cr 
 $ a_\epsilon^r $&-1.6981&-2.6679&-1.3329&-2.4935\cr 
 $ a_\epsilon^i $& 2.4997& 2.8956& 2.5274& 2.9127\cr 
 $ a_{87}^r $& 0.1923& 0.1923& 0.2025& 0.2025\cr 
 $ a_{7\epsilon}^r$&-0.7827&-1.0940&-0.8092&-1.1285\cr 
 $ a_{8\epsilon}^r$&-0.0601&-0.0819&-0.0573&-0.0783\cr 
 $ a_{87}^i $&-0.0487&-0.0487&-0.0487&-0.0487\cr 
 $ a_{7\epsilon}^i$&-0.9067&-1.0447&-0.9291&-1.0585\cr 
 $ a_{8\epsilon}^i$&-0.0661&-0.0779&-0.0637&-0.0765\cr 
\hline
\end{tabular}
\end{center}
\caption{Numerical values of the coefficients introduced in Eq.~(\ref{brnum}). We give the values
corresponding to $E_0 = (1.6 \, {\rm GeV}, m_b/20)$ and to $m_c/m_b = (0.23, 0.29)$.}
\label{tab:coeff}
\end{table}
\subsection*{Acknowledgments}
This work is supported by   the Swiss `Nationalfonds'.
W.P. is supported by the 'Erwin Schr\"odinger fellowship No. J2272' 
of the `Fonds zur F\"orderung der wissenschaftlichen Forschung' of 
Austria.

\end{document}